\documentclass{aa}

\usepackage{graphicx}

\begin{document}
   \title{Chemical compositions of four barium stars}

   \author{Y. C. Liang
	  \inst{1}\inst{,2},
	  G. Zhao
	  \inst{1},
	  Y. Q. Chen
	  \inst{1},
	  H. M. Qiu
	  \inst{1}
          \and
          B. Zhang\inst{3}\inst{,4}	  
          }

  \titlerunning{Chemical compositions of four barium stars}

   \authorrunning{Y. C. Liang et al.}
     
   \offprints{Y. C. Liang,
              email: Yanchun.Liang@obspm.fr, lyc@yac.bao.ac.cn 
          }

  \institute{National Astronomical Observatories, Chinese Academy of
          Sciences, 100012, Beijing, P. R. China
        \and
	   GEPI, Observatoire de Paris-Meudon, 92195 Meudon, France	      
         \and
           Department of Physics, Hebei Normal University, 050016, 
Shijiazhuang,
	   P. R. China          
	  \and
	    Chinese Academy of Sciences-Peking University Joint Beijing
	    Astrophysical Center, Beijing 100871, P. R. China }

   \date{Received; accepted}

   \abstract{
 We obtain abundances of 
$\alpha$, iron peak and neutron capture (n-capture) process
elements in four Ba stars HD\,26886, HD\,27271, HD\,50082 and HD\,98839
based on high resolution, high signal-to-noise spectra. 
We find that all of these Ba stars are disk stars.
Their $\alpha$ and iron peak elements are similar to the solar abundances.
The n-capture process elements are 
 overabundant relative to the Sun.
In particular, the second peak slow neutron capture process (s-process) 
elements, Ba and La,
are higher than the first peak s-process elements, Y and Zr. 
Analyzing the abundances of four sample stars,
the heavy-element abundances of the strong Ba star 
HD\,50082 are higher than those of other three mild Ba stars.
The stellar mass of
the strong Ba star HD\,50082 is 1.32M$_{\odot}$ ($+0.28$,$-0.22$M$_{\odot}$), 
which is consistent with the
average mass of strong Ba stars (1.5M$_{\odot}$).
 For mild Ba star HD\,27271, 
 we derive 1.90M$_{\odot}$ ($+0.25$,$-0.20$M$_{\odot}$), 
consistent with the
average mass of mild Ba stars (1.9M$_{\odot}$, with 0.6M$_{\odot}$ white dwarf
companion).
For mild Ba star HD\,26886, the derived 2.78M$_\odot$ 
($+0.75$,$-0.78$M$_{\odot}$)
    is consistent with
   the average 2.3M$_\odot$ of mild Ba stars 
   with 0.67M$_\odot$ companion white dwarfs within the errors. 
Mass of mild Ba star HD\,98839 is high to 3.62M$_\odot$, which inspires 
more thoughts
on the formation of Ba star phenomenon. 
Using our angular momentum conservation theoretical model
of wind accretion of Ba binary systems, 
we obtain the theoretical
heavy-element abundances of Ba stars that best fit our data.
 The results show that
the observed abundances of the typical strong Ba star HD\,50082 and 
the typical mild Ba star HD\,27271 
are consistent with the theoretical results very well. This suggests
that their heavy-element abundances 
were caused by 
accreting the ejecta of AGB stars, the progenitors of the present white dwarf
companions,
through stellar wind.
However, 
wind accretion scenario cannot explain the observed abundance pattern 
of the mild Ba star HD\,26886 with shorter orbital period ($P$=1263.2 d).
The mild Ba star HD\,98839
with high mass (up to 3.62M$_{\odot}$) 
and very long orbital period ($P>$11000 d)
may be either a star with the heavy elements enriched by itself
or a ``true Ba" star.
 

   }

   \maketitle
%

\section{Introduction}

The ``barium stars", or ``\ion {Ba} {ii}  stars" as they were first called, 
were defined as 
a class of peculiar red giants by Bidelman \& Keenan (1951). 
 These G and K giants exhibit enhanced features of 
\ion {Ba} {ii}, \ion {Sr} {ii}, CH, CN, and sometimes
C$_2$ lines. Many qualitative studies on \ion {Ba} {ii} stars have been 
developed
(e.g. Garstang 1952; Burbidge \& Burbidge 1957; Danziger 1965; 
Pilachowski 1977; Tomkin \& Lambert 1983; Smith 1984;
Kovacs 1985; Za$\check{c}$s 1994).
The results confirmed that, in addition to Ba and Sr, other 
heavy elements are also enhanced, e.g. Y, Zr, La, Ce, Pr, Nd and Sm.

Some researchers suggested that 
the overabundances of Ba stars were produced in the
interior of the \ion {Ba} {ii} star, and subsequently mixed to the surface
(e.g. Burbidge \& Burbidge 1957 for HD\,46407).
However, the absence of the unstable nucleus $^{99}$Tc
($\tau_{1 \over 2}=2\times 10^5$ yrs) 
supply evidence that they are not presently
undergoing nucleosynthesis activity and the third dredge-up process like 
asymptotic giant branch (AGB) stars.

After Burbidge et al. (1957) suggested that 
elements heavier than iron
are synthesized in the interior of AGB stars, 
it is generally believed that Ba stars
belong to binary systems and 
their heavy-element overabundances are produced by 
accreting the matter ejected by the companions
(the former AGB stars, now evolved into white dwarfs).
Many researchers have studied the binarity or heavy-element abundances of Ba 
stars 
(McClure et al. 1980; McClure 1983; McClure \& Woodsworth 1990;
Boffin \& Jorissen 1988; 
Jorissen \& Mayor 1992; Jorissen et al. 1998; Liang et al. 2000).
The mass exchange took place about $1\times 10^6$ years ago, 
so the $^{99}$Tc produced in the original thermal pulse (TP) AGB 
stars have decayed. 
The accretion may be in the forms of wind accretion,
disk accretion or common envelope ejection
(Han et al. 1995; Jorissen et al. 1998; Liang et al. 2000). 

Therefore, it is very important to combine abundance analysis
and binary characters of Ba stars 
to study their properties and formation scenarios.
 
 At present, there is a large sample of Ba stars 
 with measurements of 
orbital elements (Carquillat et al. 1998; Udry et al. 1998a, 1998b; 
Jorissen et al. 1998), absolute magnitudes and kinematics 
(G\'{o}mez et al. 1997; Mennessier et al. 1997).
However, the corresponding heavy-element abundances have not been obtained
from observations.
To explore their formation scenario,
it is necessary to study their observed abundances, combined with
their orbital elements.

Moreover, using
our angular momentum conservation 
model of wind accretion of Ba stars formation, 
and combining the AGB stars nucleosynthesis (Liang et al. 2000;
Liu et al. 2000),
we can calculate the theoretical heavy-element abundances
of Ba stars. Thus we can understand the observed abundance
patterns of the sample stars from theory.

In addition, with the high precision Hipparcos data, 
the precise photometric parameters,
improved methods to determine stellar atmospheric parameters
and developed stellar evolutionary tracks study,  
it is possible to
allow to obtain the masses and atmospheric parameters of these stars. 
For Ba stars, combining masses, element abundances and orbital parameters
provides a good understanding of their properties and formation. 
 In this paper, we discuss the properties of four Ba stars by combining
 these parameters.

This paper is organized as follows.
In Sect.\,2--4, we describe the observations and analysis methods 
in details, and
present the derived atmosphere parameters and masses.
Detailed spectral lines analysis and equivalent widths (EWs) are also 
presented.
In Sect.\,5, we show the element abundance patterns
and analyze the abundance uncertainties. In Sect.\,6, 
we calculate the theoretical
abundance trends of s-process elements in Ba stars, 
and compare them with the observed patterns of our sample
stars. Detailed discussions are given in Sect.\,7.
In Sect.\,8, we summarize this paper and describe possible future
studies.

 \section{Observations and data reduction}

The high resolution spectra of four barium stars were obtained with 
the Coud\'{e} Echelle Spectrograph and a 1024$\times$1024 TeK CCD 
attached to the 2.16m telescope at the National
Astronomical
Observatories (Xinglong, China) in 2001. The red arm of the spectrograph
with a 31.6 grooves/mm grating was used.
With a 0.5 mm slit, the resolving power is of the order
of 40 000 in the middle focus camera system.
The total wavelength coverage was 5800$-$9000 \AA$~$. 
The signal-to-noise (S/N) ratio
is greater than 150 over the whole region. 
A detailed description of technical
aspects of the spectrograph can be found in Zhao \& Li (2001).
   
The data reductions were carried out with a standard MIDAS package. 
The procedure is: 
order identification, background subtraction, flat-field correction,
order extraction, wavelength calibration, 
radial velocity shift correction, and
continuum normalization. 
Bias, dark current and scattered--light corrections 
are included in the background subtraction. 
The pixel-to-pixel sensitivity variations were corrected by using the 
flat-field.
The wavelength calibration was based on thorium-argon lamp spectra. 
In Fig.\,1, we present a portion of spectrum of the sample star HD\,98839 
in the region near \ion{Ba}{ii} $\lambda$6141.727 line.

\begin{figure}
\input epsf
\epsfverbosetrue
\epsfxsize 8.8cm  
\epsfbox{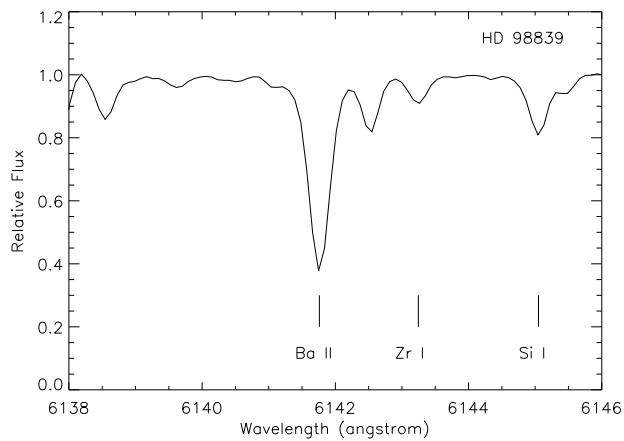}
\caption {Example of spectra of HD\,98839: \ion{Ba}{ii} $\lambda$6141.727 line, 
\ion{Zr}{i} $\lambda$6143.180 line and \ion{Si}{i} $\lambda$6145.020 line.   }
\end{figure}

The measurements of spectral line EWs were
done by applying two different methods: direct integration of the line profile
and Gaussian fitting. The latter is preferable in the case of faint lines, but
unsuitable for the strong lines in which the damping wings contribute
significantly to the equivalent width. The final EWs are weighted
averages of these two measurements, depending on the line intensity 
(see Zhao et al. 2000 for details).

{
\begin{table*} 
{ \begin{center}  \caption[] {\centering Basic data of the sample stars }  
 \begin{tabular}{ccccccccc} \hline \hline
Star name &  Sp. & V$_{\rm mag}$ & $B-V$ & $b-y$ & $\pi$  & $\sigma_\pi$& $BC$ 
& M/M$_{\odot}$     \\ 
          &      &              &      &       & (mas) &              &   &    
              	  \\\hline
HD 26886  &  G5  &7.00     &0.92 &0.596& 2.74   &  1.05       &--0.335  & 
2.78(+0.75,--0.78)      \\    
HD 27271  &  G5  &7.52     &1.00 &0.591& 6.01   &  1.13       &--0.306  & 
1.90(+0.25,--0.20)     \\
HD 50082  &  G0  &7.43     &1.03 &0.611& 4.71   &  0.99       &--0.367  & 
1.32(+0.28,--0.22)       \\ 
HD 98839  & G8II &4.99    &0.98 &0.610& 6.63   &  0.63       &--0.308  & 
3.62(+0.45,--0.00)     \\ \hline 
 \end{tabular} 
\end{center}
}
\end{table*} 
}

\section{Stellar atmospheric parameters}

Model atmospheres are specified by four parameters: effective temperature,
surface gravity, overall metallicity and microturbulent velocity.

 Effective temperature $T_{\rm eff}$ is determined from the $uvby$ system index
($b-y$) and [Fe/H]
using the empirical calibration of Alonso et al. (1999).
This calibration is based on a
large sample of field and globular cluster stars roughly covering spectral type
from F0 to K5, and derived by applying the infrared flux method.
The photometric data of sample stars are taken from Olsen (1993) and
Hauck \& Mermilliod (1998). 
The error of the photometric data is $\sigma(b-y)$=0.004 mag. 
Adopting the mean $\sigma$(Fe/H)=0.14 dex from the
spectroscopic analysis, the statistical error of $T_{\rm eff}$ is estimated to
about $\pm$50\,K. Considering a possible error of $\pm$80\,K in the
calibration, and other factors, 
we estimate the error in $T_{\rm eff}$ as high
as 100 K for our sample stars.

 The previous method to determine gravity log\,$g$ requires that
 \ion{Fe} {i} and \ion{Fe} {ii}
 lines give the same iron abundance. 
But it is well known that the derivation of iron
abundance from \ion{Fe} {i} and \ion{Fe} {ii} lines may be affected 
by many factors such as
unreliable oscillator strengths, possible 
non $local~ thermodynamic~ equilibrium$ (non-LTE) effects and uncertainties in
the temperature structure of the model atmospheres. From Hipparcos parallaxes, 
we can determine more reliable gravities. 
Using the relations:
$$
log {g \over g_{\odot}}=log {M\over M_{\odot}}
+4log {{T_{\rm eff}}\over T_{{\rm eff},\odot}}+0.4(M_{\rm bol}-M_{{\rm 
bol},\odot})
$$ 
and
$$
M_{\rm bol}=V+5+5log \pi +BC,
$$
where $M$ is the stellar mass, $M_{\rm bol}$ the absolute bolometric 
magnitude, $V$
the visual magnitude, $BC$ the bolometric correction, and $\pi$ the parallax.
We adopt solar value log\,$g_\odot$=4.44, 
$T_{{\rm eff},\odot}$=5770 K, $M_{{\rm bol},\odot}$=4.75 mag. 
The parallax $\pi$ is taken from the Hipparcos Satellite observations 
(ESA 1997). 
Stellar mass was determined from the position of the star in
$M_{\rm bol}-$log\,$T_{\rm eff}$
diagram. The corresponding errors were estimated by changing 
$T_{\rm eff}$, [Fe/H]
and $M_{\rm bol}$ according to its derivations (more details can be found 
in Chen et al. 2000). 
We adopt the stellar evolution tracks given by Girardi et al. (2000).
The bolometric correction, $BC$, is determined using the empirical calibration 
of Alonso et al. (1999). 
The uncertainty of log\,$g$ determined by this method mainly comes from the
uncertainties in the 
stellar distance, stellar mass, effective temperature and bolometric 
correction.
We estimate the error in log\,$g$ to be in the range of
0.1$-$0.3 dex for our sample stars
(Table\,2 and the Col.\,4 in Table\,5).

 The initial metallicities of the program stars were taken from the literature 
if
available. Otherwise, we obtain an initial value by judging from the spectra 
and the
color index. The final model metallicity is derived from consistency with
the other parameters in the abundance calculation.   

Microturbulent velocity $\xi _t$
was determined from the abundance analysis by requiring a zero slope of 
[Fe/H] vs. EWs. The values of the sample stars are given in Table 2.
The error of $\xi _t$ is about 0.3 km s$^{-1}$.

Some characteristic parameters of four barium stars
are presented in Table 1. The columns are: star name, spectral type, 
$V$ magnitude,
($B-V$), ($b-y$), parallax $\pi$, uncertainty
of parallax, bolometric correction and stellar mass.
The atmospheric parameters of the sample stars
are summarized in Table 2 including star name, effective temperature, 
surface gravity, microturbulent velocity and metallicity.

\begin{table} 
   \caption[] {\centering Atmospheric parameters of the sample stars} 
 \begin{tabular}{ccccc} \hline\hline
Star name & T$_{\rm eff}$ & log\,$g$      &    $\xi _t$  & [Fe/H]        \\ 
          &   (K)       &              & (km s$^{-1}$) &                 \\   
\hline
HD 26886  &    4802     & 2.22$\pm$0.3  &  1.2    &  --0.28$\pm$0.16 \\    
HD 27271  &    4874     & 2.98$\pm$0.2  &  1.4    &  --0.06$\pm$0.14       \\
HD 50082  &    4730     & 2.50$\pm$0.2  &  1.5    &  --0.40$\pm$0.14     \\ 
HD 98839  &    4866     & 2.33$\pm$0.1  &  1.6    &   +0.16$\pm$0.16           \\ \hline 
 \end{tabular} 
\end{table}

\section{Model atmospheres and spectral lines}

We adopt the model atmospheres generated by ATLAS9 code (Kurucz 1993)
to do the abundance analysis: these are LTE, 
plane-parallel, line-blanketed models.
Abundances are determined by using the input atmospheric parameters given in
Table 2 and the measured EWs. 
We usually choose the lines with EWs$=$20$\sim$120 m\AA, because the weaker 
lines
would increase random error and possibly some systematic
overestimates,
while the stronger lines are very sensitive to the microturbulence and damping.
We use the updated  oscillator strengths of spectrum lines taken from the NIST
database (http://physics.nist.gov). But
the log\,$gf$ values of n-capture process elements are taken
from the related references (Hannaford et al. 1982;
Bi$\acute{e}$mont et al. 1981; Weise \& Martin 1980;
Luck \& Bond 1991; Bi$\acute{e}$mont et al. 1982).
All of the lines used in determining abundances are given in 
Table\,3, including line and wavelength, excitation potential $\chi$,
 oscillator strengths log\,$gf$ and EWs. For \ion {Ba} {ii} $\lambda$5853.688 
line
 in HD\,98839, we find its EW is 171 m\AA, which is much lower than
 259 m\AA$~$obtained by Fern\'{a}ndez-Villaca\~{n}as et al. (1990). 
 The EW values of \ion {Ba} {ii} $\lambda\lambda$6141.727, 6496.908
 lines in this star are higher than 200 m\AA$~$obtained in this study.

\begin{figure*}
  \centering
  \includegraphics[bb=22 77 418 802,width=14.8cm,clip]{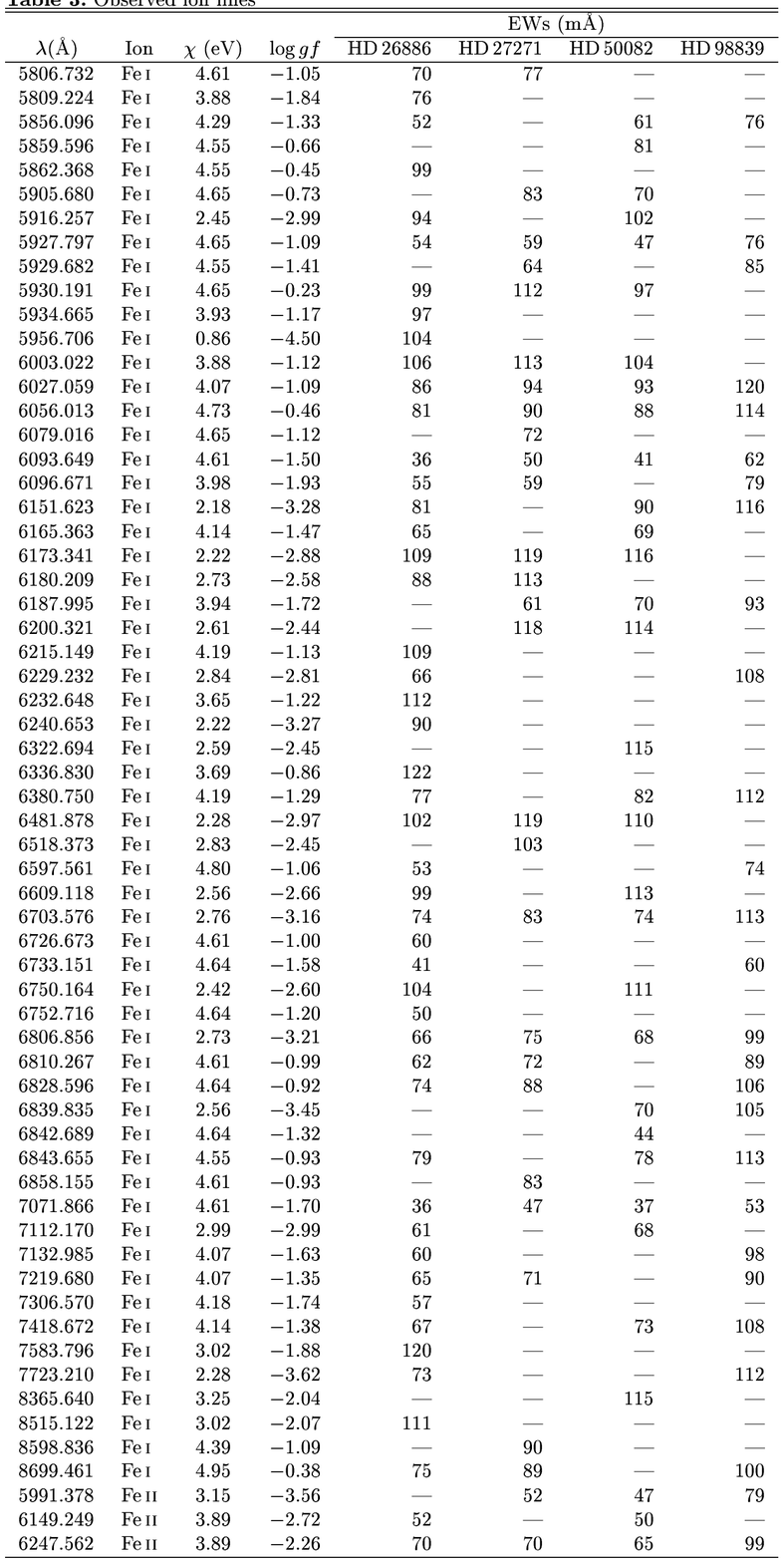}
\end{figure*}

\begin{figure*}
  \centering
  \includegraphics[bb=22 77 418 802,width=14.8cm,clip]{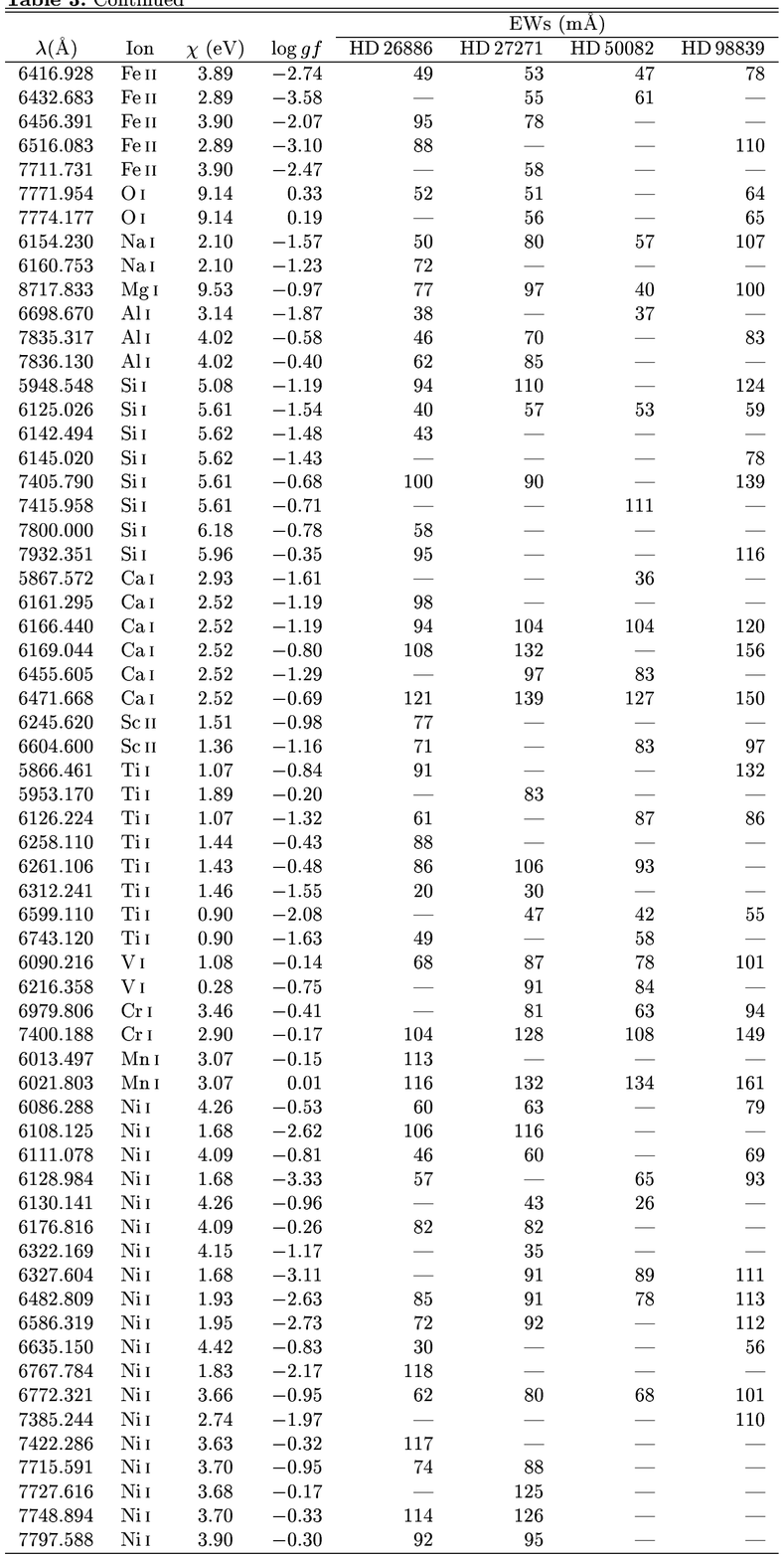}
\end{figure*}

\begin{figure*}
  \centering
  \includegraphics[bb=82 611 512 800,width=14.8cm,clip]{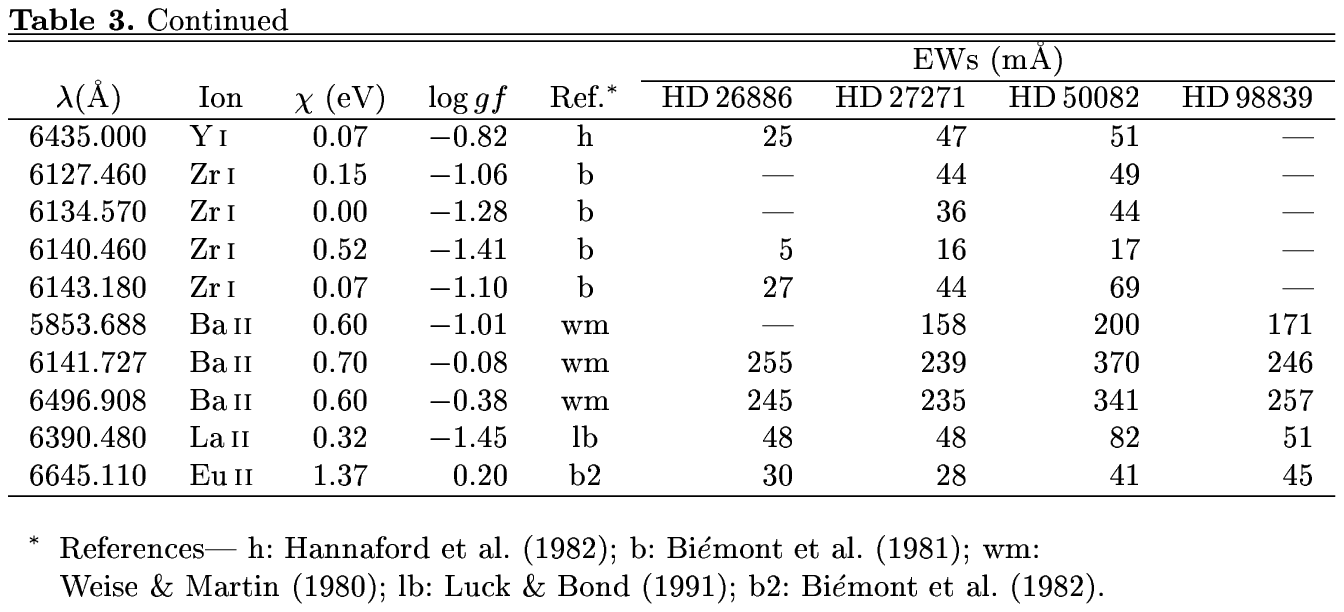}
\end{figure*}

\section{Abundances and their uncertainties}

\subsection{Abundance results}

The final abundances log\,$\epsilon$ (in the usual scale log$X$(H) = 12.0) 
and the corresponding induced [$X$/Fe] values
are presented in Table 4, where N refers to the line number used in 
calculation. 
The solar abundances are taken from Grevesse \& Sauval (1998).

Fig.\,2(a)-(d) shows the detailed abundance results of the sample stars. 
The corresponding errors
are taken from Col. 6, $\sigma_{tot}$, in Table 5. 
Detailed analysis about errors is discussed in 
next section (Sect. 5.2).  

\begin{figure*}
  \centering
  \includegraphics[bb=70 365 420 620,width=17.6cm,clip]{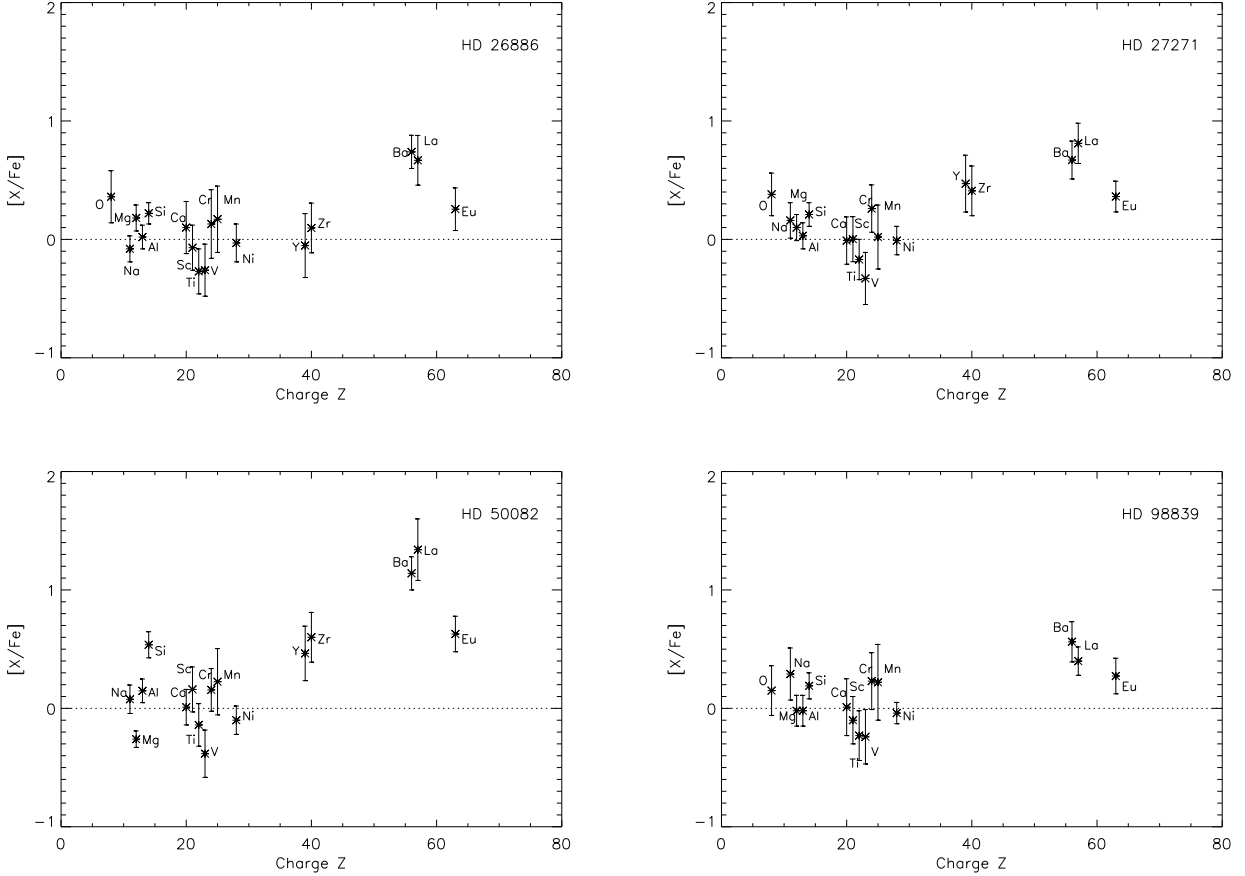}

 \caption{ 
{\bf (a)} (the top-left panel) Abundance pattern of HD26886 with the error bars  
taken from Col.6 in Table 5(1);
{\bf (b)} (the top-right panel) Abundance pattern of HD27271 with the error bars 
taken from Col.6 in Table 5(2);
{\bf (c)} (the bottom-left panel) Abundance pattern of HD50082 with the error bars  
taken from Col.6 in Table 5(3);
{\bf (d)} (the bottom-right panel) Abundance pattern of HD98839 with the error bars  
taken from Col.6 in Table 5(4).
}
\end{figure*}

\begin{figure*}
  \centering
  \includegraphics[bb=46 528 548 800,width=17.8cm,clip]{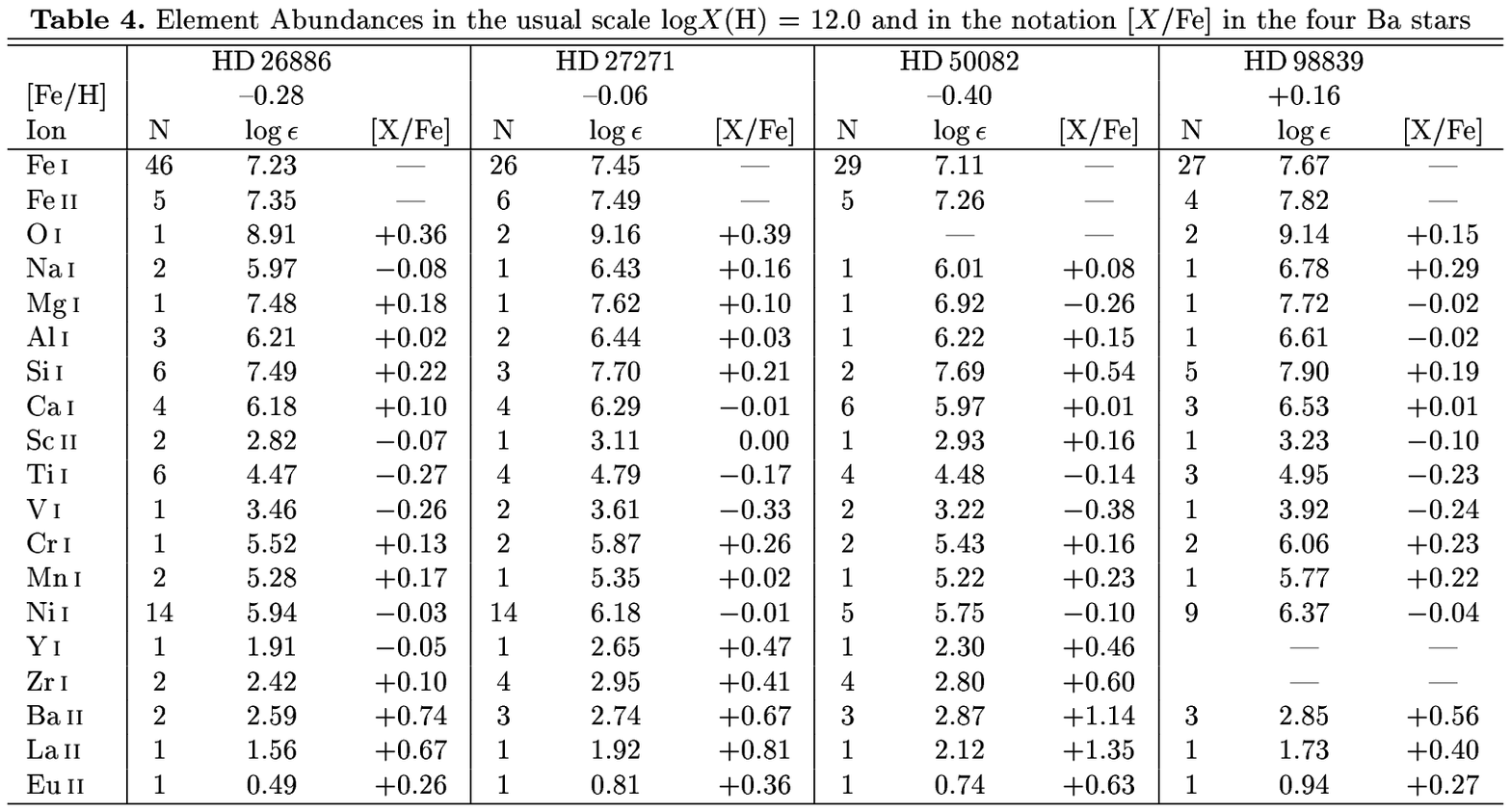}
\end{figure*}

\subsection{Errors in resulting abundances}

 Uncertainties of the abundances mainly come from equivalent widths 
 measurement, log\,$gf$ 
 values and stellar atmospheric parameters. 

Uncertainties in the EWs are set essentially by the S/N 
and the resolution of the spectra. 
An expression for the intrinsic accuracy of an equivalent width
is provided by Cayrel (1988). 
Applied to our spectra, having R=40,000 and typical S/N of 
150, we expect uncertainties in the EWs of about 2 m\AA.
We also consider the errors obtained 
by measuring EWs of spectral lines several
times in the total uncertainties.  
For an element represented
by N lines, the error is decreased by a factor $\sqrt N$. 
Thus the errors in element abundances from EWs are calculated and illustrated
on Col. 2 in Table 5, $\sigma_{EW}/\sqrt N$.
 
The errors caused by atmospheric parameters can be obtained through
changing the parameters $T_{\rm eff}$, log\,$g$, $\xi _t$ individually.
Column 3, 4, 5 in Table 5 
show the effects on the derived abundances by changing
 +100 K in $T_{\rm eff}$, 
 +0.3, +0.2 or +0.1 dex in log\,$g$, +0.3 km s$^{-1}$ in $\xi_t$.

Model metallicity uncertainties
scarcely affect the element abundances, and they can be ignored.

The uncertainty in $gf$-values of the \ion{Fe} {i} and \ion{Fe}{ii} lines 
is about 0.05 dex. 
For the elements up to Ni, the uncertainties are about 0.02$-$0.10 dex, 
and for the heavier elements, the errors are $\sim$0.20 dex (Smith et al. 1995;
Lambert et al. 1996; Junqueira \& Pereira 2001). 

The combined abundance errors caused by the above-mentioned 
uncertainties in $gf$-values, EWs-values and
stellar atmospheric parameters,
are the total errors in abundances, which are given
in the last columns in Table 5.(1)-(4) by $\sigma_{tot}$.

\begin{figure*}
  \centering
  \includegraphics[bb=42 286 556 800,width=17.8cm,clip]{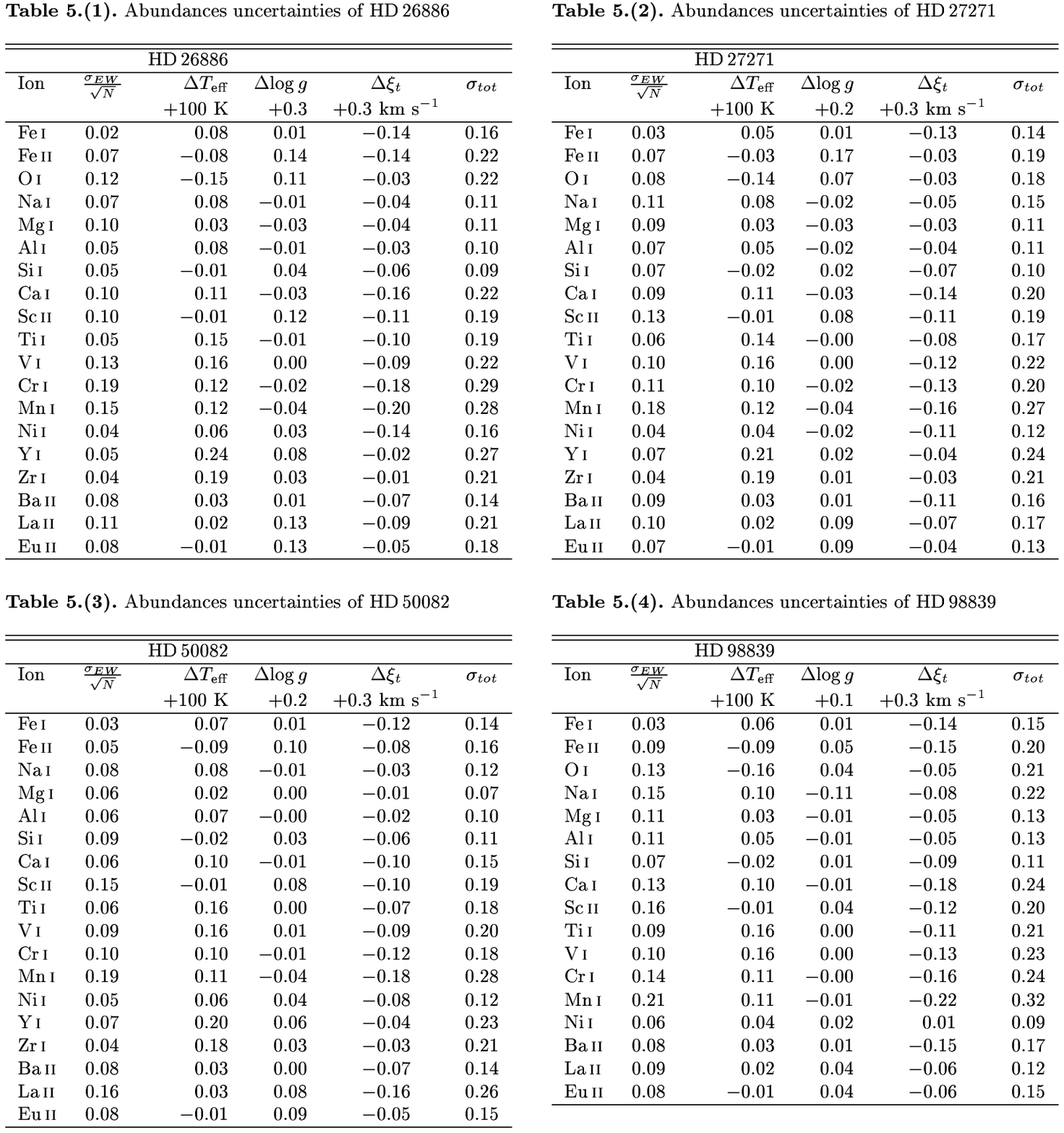}
\end{figure*}

\subsection{Abundance pattern analysis}

 Among the four sample stars, HD\,98839 shows the highest metallicity 
 with [Fe/H]=+0.16,
 which is slightly higher than solar metallicity. HD\,50082 shows the lowest 
 metallicity with [Fe/H]=--0.40. The other two stars 
 ([Fe/H] is --0.06 for HD\,27271 and --0.28 for
 HD26886) are nearer to solar metallicity
 than HD\,98839 and HD\,50082. However, 
 it is clear that all of them are disk stars.
  
 The [$\alpha$/Fe] of sample stars are similar to solar. 
 And it is clear that [$\alpha$/Fe] patterns are lower than those
 of the two halo Ba stars ([Fe/H]=--1.72, --1.43) 
 studied by Junqueira \& Pereira (2001). 
 Their two halo Ba stars show the overabundant [$\alpha$/Fe]  
 relative to solar. 
 The reason is that massive stars
 produce these $\alpha$ elements, which have been 
 ejected into interstellar medium
 with the Type II supernova explosion of the massive stars 
 since the early stage of our Galaxy.
 Most of element Fe is produced through
 Type Ia supernova explosion, which 
 occurs in the close binary system
 composed by the longer lifetime, intermediate- and low-mass stars
 (Timmes et al. 1995; Samland 1998; Liang et al. 2001).
 Thus, there is abundant element Fe in our Galaxy at its present stage.
 The [$\alpha$/Fe] patterns of our sample Ba stars also 
 show that all of these four stars are disk stars.
 In addition, it should be noticed that the [O/Fe] values of our three samples
 will be increased by 0.14 when we use the recent solar O abundance 
 log\,$\epsilon$(O)=8.69 obtained by Allende Prieto et al. (2001) 
 based on [\ion{O} {i}] $\lambda$6300 forbidden line.
 That is, the [O/Fe] ratio in HD\,26886, HD\,27271 and HD\,98839 are 0.50, 
0.53 and 0.29,
 respectively. But this increase value is still smaller than the corresponding
 abundance errors in the three sample stars: 0.22, 0.18 and 0.21, respectively.
 
 For Na and Al, these two odd elements, their abundances 
 are similar to Mg abundance,
 because these nuclei are thought to be synthesized in the same process,
 namely, carbon burning in massive stars. 
 
 The iron group elements follow Fe closely though with some scatter.
 Generally,
 Cr and Mn are slightly overabundant, and Ti, V are slightly underabundant.
 Especially, Ni, the closest element to Fe, is very similar to the Fe abundance
 in the sample stars. 
 
 However, it is not clear whether 
 the larger scatter of $\alpha$ elements 
 and iron elements in HD\,50082 
 is the natural results of its strong Ba property or not.
 
 The most important results are the obvious overabundance of 
 s-process elements ($Z>$56),
 Y, Zr and Ba, La, Eu (two peaks), in all of the four Ba stars 
 (without appropriate Y and Zr lines are used in HD\,98839
 due to spectral qualities).  
 Especially, the second peak s-process elements, Ba and La, are in the 
 highest abundances. The origin of the two peaks is the result
 of neutron magic number 50 and 82 nuclei 
 on the path of s-process nucleosynthesis
 occurred in the interiors of AGB stars (Clayton et al. 1961; 
 Seeger et al. 1965), the progenitor companions of these Ba stars. 
 There may be a third peak on element Pb
 corresponding to the neutron magic number 126. 
 But it needs high quality blue spectra to get the
\ion{Pb}{i} $\lambda$$\lambda$3683, 4057 lines to obtain the Pb abundance.
 
 To further degree, in the three sample Ba stars, 
 the first peak s-elements, Y and Zr, 
 show lower abundances than
 the second peak s-elements Ba, La etc.  This trend also was
 shown in some other Ba stars
 given by Za$\check{c}$s (1994).   
 This [hs/ls]$>$0 trend 
 (``hs" refers to the second peak elements, 
  ``ls" refers to the first peak elements) shows that 
  the neutron exposures experienced in their AGB progenitor companions 
  are strong enough to produce the more abundant ``hs" elements than
  the ``ls" elements.
  If we plot our Ba samples
  and the corresponding AGB progenitor companions on Fig.\,1 of Liang et al. 2000),
  or Fig.\,6 of Busso et al. (1995) ([ls/Fe] vs. [hs/ls]), 
  these stars should be in the [hs/ls]$>$0 regions in
  the two figures. 
  Our theoretical calculations given in next 
  section certify this point.
 
 This discrepancy between the two peak elements may be caused 
 by the different neutron exposures in the companion AGB stars.
 Namely, the higher neutron exposure in the interior of AGB stars
 benefits to the production of the heavier s-process elements
 (Liang et al. 2000; Busso et al. 1995).
 
 Comparing the abundance patterns of the two ``typical" 
 Ba star HD\,27271 and HD\,50082,
 we notice that the differences between the two peak elements (Y, Zr and Ba, 
La)
 are higher in HD\,50082 than in HD\,27271. It may be that 
 the companion AGB star of HD\,50082 experiences stronger neutron exposure
 in s-process nucleosynthesis.  
 As a result, the abundances of all of the 
 s-process elements of HD\,50082 are higher than those of HD\,27271. 
 Our theoretical calculations
 certify this analysis (see Sect. 6).
 
 As for calling HD\,50082 and HD\,27271 as ``typical" Ba stars,
 the main reason is their masses. 
 According to the scale defined by Warner (1965), there are three mild 
Ba stars
in our samples: HD\,26886, HD\,27271 and HD\,98839, and one strong Ba star
 HD\,50082.  
The average masses are 1.9M$_{\odot}$ and 1.5M$_{\odot}$ respectively
for mild Ba stars and strong Ba stars (Jorissen et al. 1998).
We obtain masses of 1.90M$_{\odot}$ and 1.32M$_{\odot}$ 
for HD\,27271 and HD\,50082 respectively, which are consistent 
with the mass judgments well. 

 HD\,26886 and HD\,98839 also show over-abundant Ba and La,
but the abundances are lower than those of HD\,50082 and HD\,27271.
The elements Y, Zr of HD\,26886 are weakly over-abundant.
These results can  
be understood basically by the weaker overabundances of heavy elements
in their AGB progenitor companions, by 
the lower neutron exposures occurred in the AGB
progenitors.

\section{Comparison with wind accretion model results}

What is the reason to cause the heavy-element overabundances of Ba stars?
It is generally believed that
the overabundances are caused by binary accretion,
namely, Ba stars accret the ejected materials from their companions, the former
AGB stars (the present white dwarfs), 
which synthesized these heavy elements by themselves and ejected
the elements into interstellar medium through stellar wind
(Liang et al. 2000 and references therein).

In our previous paper (Liang et al. 2000), 
we calculated the s-process nucleosynthesis of AGB star 
with 3M$_{\odot}$ and solar metallicity. At the same time, we set up 
an angular momentum conservation model of wind accretion in binary systems.
Using such model we calculated
the heavy-element abundances of Ba stars, and successfully 
explained the observed
abundances of some samples taken from Za$\check{c}$s (1994) and 
Busso et al. (1995).

We try to explain the abundance patterns of our 
sample stars using our theoretical method.
The calculation of the heavy element abundances of Ba stars 
is made in two separate steps. Firstly, adopting
the theory of s-process nucleosynthesis 
and the latest TP-AGB model (Straniero et al. 1995; 
Straniero et al. 1997; Gallino et al. 1998; Busso et al. 1999), 
we calculate the overabundances of the intrinsic AGB star at
each ejection. Then, combining the accreting matter predicted by the model of
wind accretion on successive occasions and mixing, 
we calculate the heavy-element
overabundances of the barium star. 
The s-process nucleosynthesis scenario and 
the change equations of orbital semi-major axis $A$ and
eccentricity $e$, $\frac{\Delta A}{A}$ and  $\frac{e\Delta e}{1-e^2}$,
were given in Liang et al. (2000) and Liu et al. (2000).

The standard case of wind accretion is:
$M_{1,0}$=3.0$M_{\odot}$,
$M_{2,0}$=1.3$M_{\odot}$, $v_{\rm ej}$=15${\rm ~km~s^{-1}}$ 
($M_{1,0}$ is the main sequence mass of the intrinsic AGB star, 
the present white dwarf, in the binary system; 
$M_{2,0}$ is the corresponding mass of the present Ba star;
$v_{\rm ej}$ is the wind velocity). 
The used accretion rate is 0.15 times of the 
Bondi-Hoyle's accretion rate (Liang et al. 2000; 
Boffin \& Za$\check{c}$s 1994).

We try to explain the observed abundances of our three Ba stars,
HD\,50082, HD\,27271 and HD\,26886, from theoretical calculations. 
Notice that there is no
eccentricity of HD\,98839 was observed, so we do not calculate
the theoretical abundances to fit its observations.
In our calculations, we try to make the calculated 
orbital period $P$ and eccentricity $e$ 
match the observations of the samples. 
Table 6 lists the observed orbital elements and their Ba classes
of the samples taken from Jorissen et al. (1998), and
their masses obtained by us (also in Table 1). 

As the studies show, 
the Ba stars with orbital period
$P>$1600 days can be formed through wind accretion
(Zhang et al. 1999; Liang et al. 2000; Liu et al. 2000).
Jorissen et al. (1998)
suggested the corresponding period is 1500 days. 
Possibly, the Ba stars with lower orbital period 
form through other scenarios: dynamically stable late case C
mass transfer or common envelope ejection. 

For our sample of Ba stars,
the orbital periods of HD\,50082 and HD\,27271 are 2896 and 1693.8
days ($>$1600 days) respectively. Their masses are consistent with  
the average masses of typical strong Ba stars (1.5M$_{\odot}$) and mild Ba 
stars (1.9M$_{\odot}$)  (see Sect. 5.3),
so we calculate the theoretical abundances of Ba stars using the standard case 
of our wind accretion model
to compare with the observed patterns.
Fig. 3(a), (b) show that the calculations (the solid lines) 
can fit observations very well within the errors for HD\,50082 and HD\,27271.  
The variable ``$a$" represents the times 
of the corresponding standard exposures in the $^{13}$C profile suggested by  
Gallino et al. (1998) (details can be found in Liang et al. 2000). 
In summary, the higher $a$ value reflects the higher neutron exposure
occurred in interiors of AGB progenitor companions.

Fig.\,3(a), (b) show that 
the required neutron exposure characteristic parameter $a$ is higher for 
HD\,50082,
$a$=2.6; while for HD\,27271, $a$=1.6 is adequate. This difference 
reflects the ``strong" and ``mild" Ba properties of each star.

For HD\,26886, since its mass is 2.78M$_{\odot}$ (though with large error),
we adopt 2.5M$_{\odot}$ as the initial mass of a Ba star for a binary system
in our calculation, instead of 1.3M$_{\odot}$ in the standard case. 
However, the fit failed (Fig. 3(c), the bottom panel).
Maybe this is due to 
the lower orbital period of HD\,26886 (1263.2 days) than 
the lower limit of wind accretion scenario (1600 or 1500 days). 
Perhaps dynamically stable late case C
mass transfer or common envelope ejection scenario
is the formation source of HD\,26886. 
However, more studies are needed to understand this suggestion.

\begin{figure}
  \centering
  \includegraphics[bb=179 710 430 795,width=9.8cm,clip]{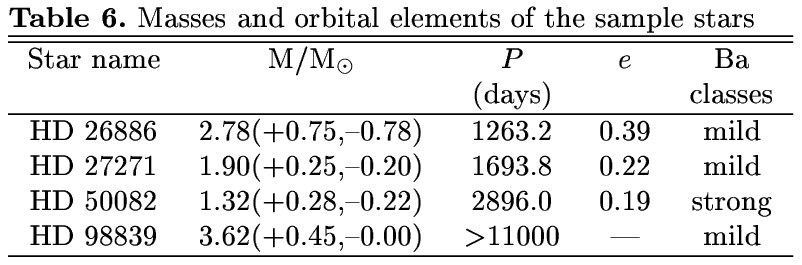}
\end{figure}

\begin{figure}
  \centering
  \includegraphics[bb=77 307 248 678,width=8.8cm,clip]{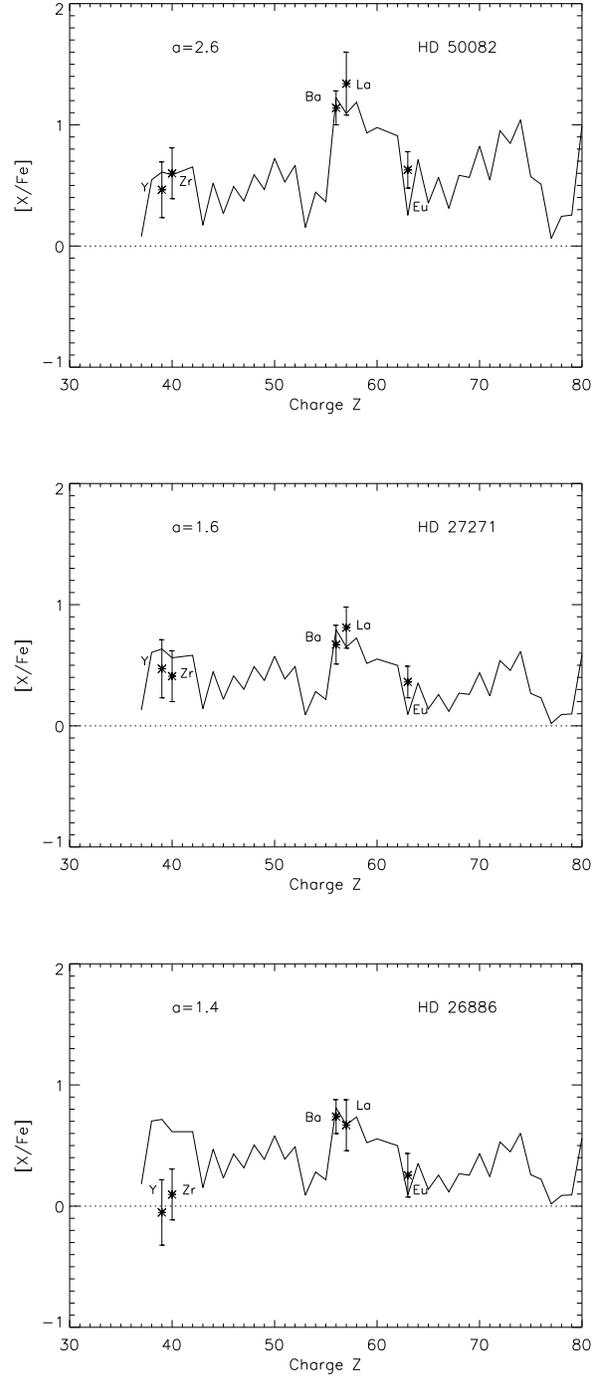}

 \caption{ The fittings of the predicted to observed heavy 
element abundances of barium stars using our wind accretion models. 
{\bf (a)} for HD\,50082 (the top panel); 
{\bf (b)} for HD\,27271 (the middle panel);
{\bf (c)} for HD\,26886 (the bottom panel).
}
\end{figure}

\section{Discussions}
  \subsection{Stellar masses and mild Ba stars, strong Ba stars}
  Adopting stellar evolution tracks given by Girardi et al. (2000) and some
  related parameters (see Sect. 3), we calculate the masses of the sample Ba stars
   (see Table. 1, 6).  
  Their masses are consistent with the statistical and theoretical
   results. Table 9 of Jorissen et al. (1998) shows that
   the average mass of mild Ba stars is 1.9M$_\odot$ with
   0.60M$_\odot$ companion white dwarfs. For HD\,27271, a
   mild Ba star, its mass is 1.90M$_\odot$, which fits  
   the average 1.9M$_\odot$ of mild Ba stars very well.
   As for HD\,26886, the calculated 2.78M$_\odot$
    is consistent with
   the average 2.3M$_\odot$ of mild Ba stars 
   with 0.67M$_\odot$ companion white dwarfs (Jorissen et al. 1998)
   within the errors. 
   HD\,98839 is discussed in next section.
   
    For the strong Ba star HD\,50082, we
   obtain a mass of 1.32 M$_\odot$, 
   which is consistent with the average mass 1.5M$_\odot$ of strong Ba stars
   with 0.60M$_\odot$ companion white dwarfs given by Jorissen et al. (1998). 

  Besides the different masses between the strong and mild Ba stars,
  their abundance patterns are different too (Sect. 5.3).
  What causes the discrepancy of abundances in these two Ba groups?
  Jorissen et al. (1998) carefully discuss the possible reasons.
  Considering our results, we emphasize and analyze 
  two possible reasons again here:
  (i) a lower mass for the barium star results in a smaller dilution
  of the accreting materials in envelope of the Ba star; 
 (ii) the lower 
  metallicity results in larger heavy-element overabundances
  in the AGB progenitor.
  Our results are consistent with these two points.
  The strong Ba star
  HD\,50082 has the lowest mass and lowest metallicity in the four samples.
  Thus it shows the highest heavy-element overabundances.

   \subsection{Inspiration from HD\,98839: a ``false Ba" star?}

  For HD\,98839, the atmosphere parameters,
  [Fe/H] and over-rich Ba abundance obtained by us are similar to those of 
  other researchers (Pilachowski 1977;
  Fern\'{a}ndez-Villaca\~{n}as et al. 1990; Mishenina 1996).
  We estimate its mass to be 3.62M$_\odot$.  
  Its high mass and heavy-element
  overabundance pose
  some questions
   on Ba stars formation through binary accretion and also
   on nucleosynthesis scenario of AGB stars.
   
   The mass of HD\,98839 is 3.62M$_\odot$, higher than
   the average masses of mild Ba stars (1.9 or 2.3M$_\odot$, see Sect.\,7.1). 
   And there is also a discrepancy between this value and the suggestion
   of Han et al. (1995), in which the authors assigned masses of less than
   3M$_\odot$ for all Ba stars.  
   
    How can these results and discrepancies be understood?
   On one hand, this mild Ba star may have been classified erroneously
   as Ba star, namely it is ``false Ba" star, and the real story is that 
   it produced the overabundant heavy elements by itself. 
   However, according to the derived parameters, 
   $M_{\rm bol}$=$-$1.2 and $T_{\rm eff}$=4866\,K,    
   HD\,98839 should not evolve to AGB stage.  
   Generally, AGB stars have $M_{\rm bol}$ from $-$1.7 to $-$5.5 and $T_{\rm 
eff}$
   from 3200\,K to 3850\,K respectively 
   (Smith \& Lambert 1985, 1986, 1990).
   In this case, 
   if the heavy element overabundances of HD\,98839 is an intrinsic
   nucleosynthesis event within itself, the
   over-rich s-process elements is usually associated with 
   mixing process at the helium core flare, which    
   is consistent with the alternative origin for the \ion {Ba} {ii}
   star phenomenon    
   suggested  
   by Malaney (1987) and Malaney \& Lambert (1988).
   Also,   
   Mennessier et al. (1997) 
   suggested that some of their mild Ba stars
   with mass higher than 3M$_\odot$ may be lower luminosity, post He-flare 
stars.

   Moreover, HD\,98839 displays very long orbital period ($>$11000 days)
   up to the upper detected limit (Jorissen et al. 1998),
   and there is no observed binary orbital eccentricity. 
   Thus, it is not unrealistic to consider it
   as an independent star.  
 
   However, if these high mass mild Ba stars are 
   really ``true Ba" stars,
   and the binary accretions cause their heavy element
   overabundances, it will challenge the present nucleosynthesis scenario of
   AGB stars. At present, the popular viewpoint  
   is that 1$-$3M$_\odot$ low mass AGB stars with $^{13}$C neutron source are 
   the main nucleosynthesis sites of s-process elements, 
   while the intermediate
   mass AGB stars with $^{22}$Ne neutron source cannot fit to the observations 
   (Malaney 1987; Busso et al. 1995; 
   Busso et al. 1999; Liang et al. 2000). 
   However, if masses of Ba stars are higher than
   3M$_\odot$, their companions should be intermediate mass stars on the
   main-sequence. How the intermediate mass AGB stars result
   in the heavy element overabundance of these Ba stars? 
   If this is the case, what kind of neutron exposure scenario works? Single 
exposure?
   More studies are needed on s-process nucleosynthesis scenario
   of AGB stars. Certainly, we need also the large sample of high mass mild
   Ba stars to check its abundance patterns.
      
   In addition to the high mass (3.62M$_\odot$) of mild Ba star, HD\,98839,
   obtained by us, 
   there are also some other high mass samples. 
   Smith et al. (1980) calculated
   mass of $\zeta$\,Cap is about 3.5M$_\odot$, but with large
   error (+3.6,--1.8M$_{\odot}$) (also see Boffin \& Jorissen 1988). 
   Mennessier et al. (1997) studied higher mass Ba stars.
   They estimated that the masses are up to 7M$_\odot$ for their
   group S and 4.5M$_\odot$ for group C in which there are mild Ba stars.
   Monte Carlo simulations of Karakas et al. (2000)
   predict that the masses of Ba stars may be up to 6M$_\odot$.
       
   Therefore, it is necessary to pay more attention on
   these high mass mild Ba
   stars.

\section{Summary}
   
We obtain abundances of $\alpha$, iron peak and s-process elements 
in four Ba stars, HD\,50082, HD\,27271, HD\,26886 and HD\,98839, 
 using high resolution and high S/N spectra.
The results show that all of samples are disk stars.
They show the 
similar metallicities and the similar abundances of $\alpha$-elments 
and iron peak elements to the solar values.
All of them show s-process element overabundances relative the solar.
We obtain their masses using the highly precise Hipparcos
data, precise photometric parameters and stellar evolution tracks.
The mass results fit well the average characteristic masses of strong and mild 
Ba stars. And the high mass of mild Ba star HD\,98839 pose more 
thoughts on Ba star phenomenon.

Using our wind accretion model of binary systems and
AGB stars s-element nucleosynthesis calculation, we conclude that the
wind accretion formation scenario is possible for HD\,50082 and HD\,27271.
Wind accretion cannot explain the observations of HD\,26886 with 
the short orbital
period ($P<$1600 days). The mild Ba star HD\,98839
with high mass and very long orbital period 
may be either a ``false Ba" star with the heavy elements enriched by itself
or a ``true Ba" star.

Fortunately, though we only have four samples,
they represent wide Ba star groups.  
From stellar masses,
there is one strong Ba star among them,
and also mild Ba stars; there are the members with typical mass of Ba stars,
also there are the high mass mild Ba stars, which are controversial.
From orbital elements, 
there are samples in typical wind accretion orbital period range,
there is one sample with very long orbital period, up to the detected limit,
and there is one sample with shorter value,
which may correspond to other formation scenarios. 
So these samples are very interesting in understanding 
the abundance patterns and formation scenarios of Ba stars.
 
Certainly, we need larger samples to understand the nature of Ba stars.
At present, there is a large sample of Ba stars with 
orbital elements (Carquillat et al. 1998; 
Udry et al. 1998a, 1998b; Jorissen et al. 1998),
but without the corresponding element abundances.
It is very interesting and necessary to obtain their element abundances
to be combined with the orbital elements.
It is specially important to compare the related results of the two groups
with different orbital periods: $P<$1600 and $P>$1600 days, and
to study the stars with very long orbital periods.
These results will help us to 
understand the various formation scenarios of Ba stars.
At the same time, 
it is important to obtain their masses,
particularly, for the possible high mass mild Ba stars. 
Maybe these high mass mild Ba stars can
give more clues on AGB stars nucleosynthesis. 
Moreover, it will be helpful
to understand the different abundance patterns between
``strong" and ``mild" Ba stars.

There is a larger sample of Ba stars (318 from L$\ddot{u}$ catalogue 1991) 
have been studied whose 
absolute magnitudes and kinematics by G\'{o}mez et al. (1997).
Their results show that the samples of Ba stars are an inhomogeneous group,
three of them among the four groups 
are disk population stars, and the fourth group contains halo stars. 
 Mennessier et al. (1997) confirmed further this inhomogenity.
If we can supplement the heavy-element abundances to 
compare the various characters of 
these different groups, it will be a stronger method to understand 
the natural properties of Ba stars. 

In addition, it is better to obtain the carbon and oxygen abundances 
together with s-process elements. 
The reason is that AGB stars are important nucleosynthesis sources of element 
carbon, 
which is taken
out from stellar interiors together with s-process elements by the 
third dredge-up process (Liang et al. 2000; Liang et al. 2001).
Element carbon will be then accreted by Ba stars, 
the companions of binary systems.

Considering Ba stars from stellar population,
it is interesting to combine Ba stars with CH stars.
Some reseachers suggest that metal-rich stars evolve into Ba stars
and the metal-poor stars evolve to CH stars, which 
are metal-poor stars (Vanture 1992a, 1992b).
At present, it is believable that 
nucleosynthesis of metal-poor AGB stars are more efficient than
the metal-rich one (Busso et al. 1999; Busso et al. 2001). 
Thus, it is valuable to study the 
heavy-element abundances of metal-poor CH stars.
But CH stars are fainter, some are as low as 12 mag, which 
makes more difficult
to get the required spectra.
 
CH subgiants (Luck \& Bond 1982, 1991; 
Smith et al. 1993),
Ba dwarfs (North et al. 1994) and extrinsic S stars (Jorissen et al. 1998)
should be the closest groups to Ba giants. It will be interesting
to combine and compare all of these group stars.
Perhaps we will then be able to study Ba stars in an evolutionary sequence.

 \begin{acknowledgements}
We thank the anonymous referee for very useful suggestions on the original
manuscript.
 We thank Dr. Rafael Guzm\'{a}n for his great help
 for improving the English expression. Y. C. Liang thanks Dr. Francois Hammer
 and other staffs of GEPI for their friendly help.
  Y. C. Liang thanks Huawei Zhang, Jianrong Shi, Zhenxi Zhu, Guangshun Bai 
and Gang Liu for their help with data reduction.
This research work is supported by the National Natural Science Foundation of
China under grant No. 10173014 and NKBRSF G1999075406. 
\end{acknowledgements}

\end{document}